\documentstyle[aps,twocolumn,epsfig]{revtex}
\begin{document}
\newcommand{\ve}[1]{\mbox{\boldmath $#1$}}
\twocolumn[\hsize\textwidth\columnwidth\hsize
\csname@twocolumnfalse%
\endcsname

\draft

\title{Bose-Einstein condensates with attractive interactions on a ring}
\author{G. M. Kavoulakis}
\date{\today}
\address{Mathematical Physics, Lund Institute of Technology, P.O. Box 118,
        S-22100 Lund, Sweden}

\maketitle

\begin{abstract}

Considering an effectively attractive quasi-one-dimensional Bose-Einstein
condensate of atoms confined in a toroidal trap, we find that the system 
undergoes a phase transition from a uniform to a localized state, as the 
magnitude of the coupling constant increases. Both the mean-field 
approximation, as well as a diagonalization scheme are used to attack the
problem.

\end{abstract}

\vskip0.0pc]

Effectively attractive Bose-Einstein condensates have concentrated a lot
of attention recently, in connection with the experimental formation of
bright solitons in them \cite{sol1,sol2}. More specifically, in the
experiments of Refs.\,\cite{sol1,sol2}, $^{7}$Li atoms were confined in
quasi-one-dimensional traps. With use of the Feschbach resonances 
\cite{Ket} the effective coupling constant that describes the atom interactions 
was then tuned and as it became negative -- corresponding to an effective  
attraction between the atoms -- localized states, ``bright soliton trains" 
were observed to form in Ref.\,\cite{sol1}, while a single bright soliton
was observed in Ref.\,\cite{sol2}. References \cite{refs,Ueda} have examined
theoretically these systems. In the limit where transversely to the 
long axis of the trap the gas is in the lowest harmonic-oscillator level, the 
transverse degrees of freedom are frozen out and the system is essentially
one-dimensional \cite{Kettnew}. 

Motivated by these developments, we study here an effectively attractive
one-dimensional Bose-Einstein condensate of atoms confined in a toroidal trap.
In a recent theoretical paper Kanamoto, Saito, and Ueda have investigated
the ground state and the low-lying excited states of such a system \cite{Ueda}
(see also Ref.\,\cite{CCR} for a detailed discussion of this problem). 
As shown, the Gross-Pitaevskii mean-field theory 
predicts a quantum phase transition between a uniform state and a localized 
state as the absolute value of the strength of the interaction 
inreases. Furthermore, numerical diagonalization of the
Hamiltonian for a finite number of bosons shows that the transition 
in this case is smeared out, as expected in finite systems.

In our study we examine the same problem using different techniques.
Initially we use the mean-field approximation with a properly
chosen variational wavefunction to study the phase transition and the
order parameter in the two phases. Then, working in the same truncated
space we use a Bogoliubov transformation to diagonalize the Hamiltonian.
Having diagonalized the problem, we examine the lowest state of the system, 
the low-lying excited states, as well as the depletion of the condensate
at the region of the transition. Our results are consistent
with those of Ref.\,\cite{Ueda}.

Let us therefore consider a Bose-Einstein condensate in a toroidal trap.
Following Ref.\,\cite{Ueda}, we assume that the system contains $N$ bosons, 
that the radius of the torus is $R$ and its cross section is $S = \pi r^2$,
with $r \ll R$. If $\hat{\psi}(\theta)$ is the field operator, the Hamiltonian 
is
\begin{equation}
\hat{H} = \int_0^{2\pi} d\theta
       \left[-\hat{\psi}^{\dagger}(\theta)\frac{\partial^2}{\partial \theta^2}
       \hat{\psi}(\theta)
    +\frac{U_0}{2}\hat{\psi}^{\dagger}(\theta)\hat{\psi}^{\dagger}(\theta)
    \hat{\psi}(\theta)\hat{\psi}(\theta)\right],
\label{Ham}
\end{equation}
where $U_0 = 8\pi aR/S$, with $a$ being the scattering length for elastic
atom-atom collisions, and $\theta$ is the azimuthal angle. Here the length 
is measured in units of $R$ and the energy in units of $\hbar^2/2mR^2$, with 
$m$ being the atom mass.

Let us start with the mean-field approach. Within this approximation
the system is described by a single wavefunction, the order parameter
$\psi({\bf r})$, and the many-body state is the product $\Pi_i \psi({\bf r}_i)$,
with $i = 1, \dots, N$, where $N$ is the number of atoms. Therefore this 
approximation ignores correlations between the atoms and in general it has 
a higher energy than the exact solution that one can get by diagonalizing 
the Hamiltonian.

In this problem it is natural to work in the basis of plane-wave states 
$\phi_l(\theta) = e^{il\theta}/\sqrt{2\pi}$ and according to the
analysis of Ref.\,\cite{Ueda}, the order parameter close to the transition
consists of the state $\phi_0$ (the dominant component), and the states
$\phi_{\pm 1}$. To get a simple physical picture, we thus develop a
variational approach, expanding $\psi(\theta)$ in the basis of the
$\phi_{l}$ states and keeping only these three components. This is a reasonable
assumption, since states with higher values of $l$ have higher kinetic energy, 
and indeed as shown in Ref.\,\cite{Ueda} this approximation gives qualitatively 
(but not quantitatively) good results. We thus write
\begin{eqnarray}
  \psi(\theta) = c_{-1} \phi_{-1} + c_{0} \phi_{0} + c_{1} \phi_{1}.
\label{wf}
\end{eqnarray}
Since the dominant component of $\psi$ is $\phi_0$, therefore 
$|c_{0}| \gg |c_{-1}|$, and $|c_{0}| \gg |c_{1}|$. Because of the symmetry
of the problem, $|c_{-1}| = |c_{1}|$, which also guarantees that the total
angular momentum of $\psi$ is zero, as it should. The normalization condition
imposes the further constraint $|c_{-1}|^2 + |c_{0}|^2 + |c_{1}|^2 =
1$. 

To proceed, we express the energy per particle ${\epsilon_0}$ in terms of
the coefficients $c_i$, which are then determined by minimizing ${\epsilon_0}$
with respect to them \cite{KMP},
\begin{eqnarray}
    {\epsilon_0} = 2 |c_{1}|^2 + \frac {\gamma} 2 [|c_{0}|^4 + |c_{-1}|^4 +
   |c_1^4| + 4 |c_{0}|^2 |c_{-1}|^2 
\nonumber \\
   + 4 |c_0|^2 |c_1|^2 + 4 |c_{-1}|^2 |c_1|^2 + 2 c_{0}^2 c_{-1}^* c_1^*
                          + 2 (c_{0}^2)^* c_{-1} c_1],
\label{eney}
\end{eqnarray}
where $\gamma = N U_0 / (2 \pi)=4 N a R/S$ is essentially the ratio between the
interaction energy and the kinetic energy. If $c_j = |c_j| e^{i \theta_j}$, then
the last two terms in Eq.\,(\ref{eney}) are equal to $4 |c_0|^2 |c_1|^2
\cos(\theta_{-1} + \theta_1 - 2 \theta_0)$. To minimize ${\epsilon_0}$ one has
to choose $\theta_{-1} + \theta_1 - 2 \theta_0 = 0$.  Since $|c_{0}|^2 = 1 - 2
|c_{1}|^2$,
\begin{eqnarray}
        {\epsilon_0} - \frac {\gamma} 2 = 2 |c_{1}|^2 (1 + 2 \gamma)
	 - 7 \gamma |c_{1}|^4.
\end{eqnarray}
Here $\gamma/2$ is the interaction energy per particle of the uniform 
state. The above equation implies that for $\gamma > \gamma_{\rm cr}$,
where $\gamma_{\rm cr} = -1/2$, the minimum of the energy occurs for
$c_0 = 1$, and $c_{-1} = c_1= 0$. In this regime the density of the system is
uniform and the energy per particle is $\gamma/2$. On the other hand, if 
$\gamma < \gamma_{\rm cr}$, $\epsilon_0(|c_{1}|^2)$ is like a mexican hat, and  
its minimum occurs for all three $c_{i} \neq 0$, the energy per particle is 
lower than $\gamma/2$ and the cloud develops a non-uniform density.
Our approach implies that the transition takes place for the same critical
value of $\gamma$ as in the exact solution of the mean-field approach
\cite{Ueda,CCR}. It is interesting that within our scheme including more basis 
states in the order parameter does not affect $\gamma_{\rm cr}$. More 
precisely, including the states with $|l| \le m$, the quadratic terms 
in the energy have the form
\begin{eqnarray}
        {\epsilon_0} - \frac {\gamma} 2 = 2 |c_{1}|^2 (1 + 2 \gamma)
	     + 2 |c_{2}|^2 (4 &+& 2 \gamma) 
\nonumber \\
	     + \dots + 2 |c_{m}|^2 (m^2 &+& 2 \gamma).
\end{eqnarray}
From the above expression it is obvious that $\gamma_{\rm cr}$ is always
$-1/2$, essentially because of the high kinetic energy of atoms in states
with high values of $l$, that scales as $l^2$. 

Let us now examine the non-uniform state when $\gamma \alt \gamma_{\rm cr}$. 
Demanding that $\partial {\epsilon_0}/\partial |c_1|^2 = 0$, we get  
\begin{eqnarray}
|c_{\pm 1}|^2 = \frac 2 {7 \gamma} (\gamma - \gamma_{\rm cr});
|c_0|^2 = 1 - \frac 4 {7 \gamma} (\gamma - \gamma_{\rm cr}).
\label{cf}
\end{eqnarray}
Figure 1 shows the corresponding order parameter. In addition 
\begin{eqnarray}
   {\epsilon_0} - \frac {\gamma} 2 = 
  \frac 4 {7 \gamma} (\gamma - \gamma_{\rm cr})^2. 
\end{eqnarray}
If $\gamma = -1/2 - \delta$, where $\delta \to 0$ and $\delta > 0$,
the above equation implies that $\epsilon_0 = -1/4 - \delta/2 - 8 \delta^2/7$,
which should be compared with the exact mean-field solution \cite{Ueda},
$\epsilon_0^{\rm ex} \approx -1/4 - \delta/2 - 2 \delta^2$.
Therefore our trial wavefunction reproduces the linear correction, 
but it fails in the quadratic term and it has higher energy. This is not a 
surprise, however, since our approach is variational and it does not 
necessarily capture all these features. Including more basis states,
one can improve the wavefunction and the energy to the desired order
in $\delta$. 

Turning to our alternative approach of the problem, where we will go beyond
the mean-field approximation 
\begin{figure}
\begin{center}
\epsfig{file=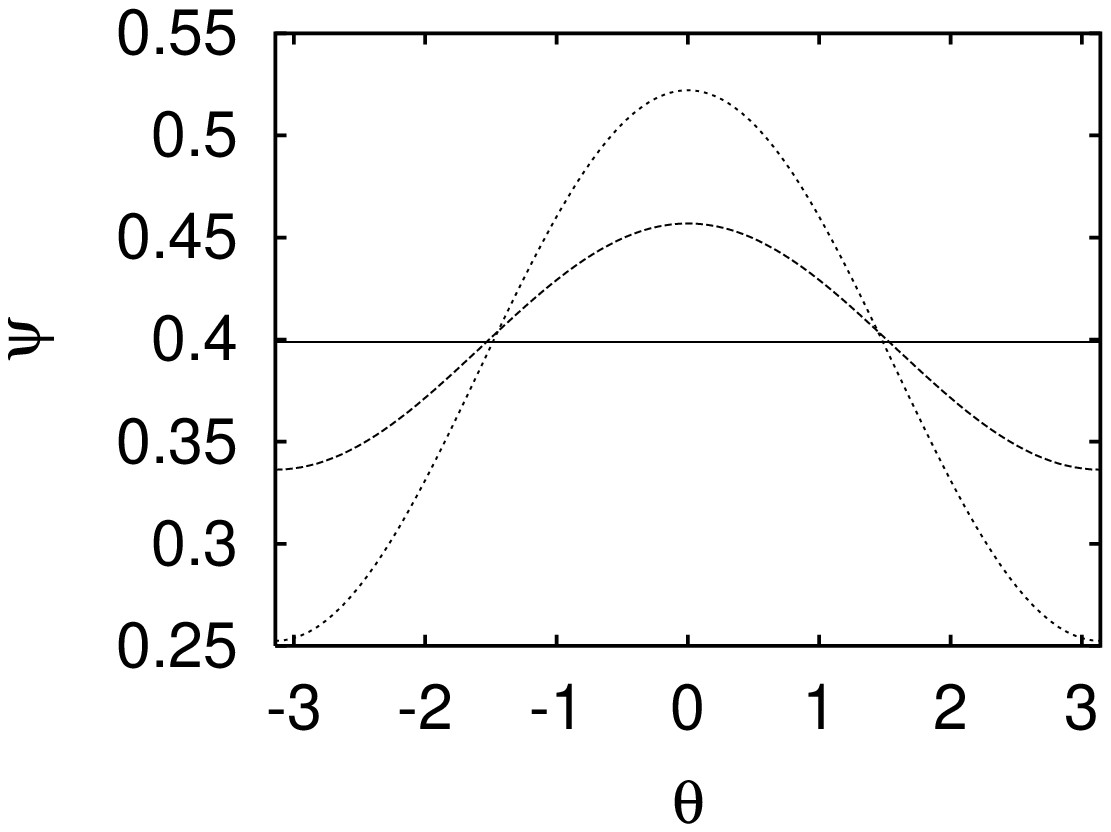,width=8.0cm,height=6.0cm,angle=0}
\vskip0.5pc
\begin{caption}
{The magnitude of the trial order parameter $\psi$, Eq.\,(\ref{wf}), calculated
within the mean-field approach, with the coefficients $|c_i|$ given by
Eq.\,(\ref{cf}). The phases of $c_{i}$ are chosen to be equal. One phase
is arbitrary, the other one is chosen so that the energy is minimum, and
the third one is free (reflecting the rotational invariance of the problem)
and it has been chosen so that the maximum of the
wavefunction is at $\theta = 0$. The straight line corresponds to a
uniform condensate, with $\gamma > \gamma_{\rm cr}$, the less localized
state to $\gamma = -0.51$, and the more localized state to $\gamma = -0.55$.}
\end{caption}
\end{center}
\label{FIG1}
\end{figure}
\noindent
and we will calculate terms in the energy 
of order $1/N$, let us expand the field operator $\hat{\psi}$,
\begin{eqnarray}
\hat{\psi}(\theta) = \phi_{-1} \, \hat{c}_{-1}
		 + \phi_{0} \, \hat{c}_0
		 +\phi_{1} \, \hat{c}_1,
\label{exph}
\end{eqnarray}
where $\hat{c}_l$ is now the annihilation operator of an atom with angular 
momentum $l$. Again we have restricted ourselves 
to the states with $l=0, \pm
1$. In this basis the Hamiltonian (\ref{Ham}) can be written as
\begin{eqnarray}
\hat{H}= \sum_{l} l^2 \hat{c}_l^{\dagger} \hat{c}_l
+ \frac{U_0}{4\pi} \sum_{klmn} \hat{c}^{\dagger}_k
\hat{c}_l^{\dagger} \hat{c}_m \hat{c}_n \, \delta_{m+n-k-l}.
\label{Haml}
\end{eqnarray}
A similar Hamiltonian has been studied in Ref.\,\cite{KMP} in the context
of weakly-interacting Bose-Einstein condensates under rotation. Denoting
the basis vectors as
\begin{eqnarray}
|m \rangle = |N_{-1},N_0,N_1 \rangle \equiv
|(-1)^m, 0^{N-2m}, (+1)^m \rangle,
\label{mu}
\end{eqnarray}
where $N_l$ is the occupancy of the state $\phi_l$, with $\sum_l N_l = N$, and
$\sum_l l \, N_l=0$, it is straightforward to calculate the diagonal matrix 
elements,
\begin{eqnarray}
\langle m | \hat{H} | m \rangle = 2 m
+ \frac {U_0} {4 \pi} [(N-2m)(N-2m-1) 
\nonumber \\
+ 2 m (m-1) + 8 m (N-2m) + 4 m^2].
\end{eqnarray}
Assuming that the system is close to the transition point, but on the 
side of the ``uniform state" (in the infinite-$N$ limit of the mean-field 
approximation), then $m$ is of order unity and thus $m \ll N$. In this limit,
\begin{eqnarray}
\langle m | \hat{H} | m \rangle \approx  \frac \gamma 2 (N-1) 
+ 2 m (1 + \gamma),
\label{meappr1}
\end{eqnarray}
where $\gamma (N-1)/2$ is the total energy of a condensate with uniform density.
In Eq.\,(\ref{meappr1}) terms of order $m^2$ have been neglected, and therefore
this approach is not valid for $\gamma < \gamma_{\rm cr}$ (since $m$ becomes 
of order $N$ there).

Turning to the off-diagonal matrix elements,
\begin{equation}
\langle m | \hat{H} | m+1 \rangle =
     \frac {U_0} {4 \pi} 2 \sqrt{(N-2m) (N-2m-1) (m+1)^2},
\label{meappr}
\end{equation}
and in the limit $m \ll N$,
\begin{eqnarray}
\langle m | \hat{H} | m+1 \rangle = \gamma (m+1).
\label{meappr2}
\end{eqnarray}
From Eqs.\,(\ref{meappr1}) and (\ref{meappr2}) $\hat{H}$
can be written as
\begin{eqnarray}
\hat{H} - \frac \gamma 2 (N-1) =  \phantom{XXXXXXXXXXXXXXXX}
\nonumber \\
 (1 + \gamma) (\hat{c}_{-1}^{\dagger} \hat{c}_{-1} +
		\hat{c}_{1}^{\dagger} \hat{c}_{1})  +
   \gamma      (\hat{c}_{-1}^{\dagger} \hat{c}_{1}^{\dagger} +
		\hat{c}_{-1} \hat{c}_{1}).
\label{hamapl}
\end{eqnarray}
The above Hamiltonian can be diagonalized with use of a Bogoliubov
transformation \cite{Bog},
\begin{eqnarray}
\hat{b} &=& \lambda_1 \hat{c}_{-1}^{\dagger} + \lambda_2 \hat{c}_{1}
\nonumber \\
\hat{d} &=& \lambda_2 \hat{c}_{-1} + \lambda_1 \hat{c}_{1}^{\dagger}.
\end{eqnarray}
For $\hat{b}$ and $\hat{d}$ to satisfy bosonic commutation relations,
$\lambda_2^2 - \lambda_1^2 = 1$. Following the usual tricks, $\hat{H}$
can be written in the diagonal form
\begin{eqnarray}
 \hat{H} - \frac \gamma 2 (N-1) = E_0
  + E \, (\hat{b}^{\dagger} \hat{b} + \hat{d}^{\dagger} \hat{d}),
\label{hamapld}
\end{eqnarray}
where
\begin{eqnarray}
E_0(\gamma) = 2 [(1+\gamma) \lambda_1^2 -
     \gamma \lambda_1 \lambda_2],
\end{eqnarray}
and
\begin{eqnarray}
E(\gamma) = (1+\gamma) (\lambda_1^2 + \lambda_2^2) -
    2 \gamma \lambda_1 \lambda_2.
\end{eqnarray}
In addition, to eliminate the off-diagonal terms one has
\begin{eqnarray}
\frac {2 \lambda_1 \lambda_2}
    {\lambda_1^2  + \lambda_2^2} = \frac {\gamma} {\gamma+1}.
\label{con}
\end{eqnarray}
Parametrizing, i.e, writing $\lambda_1 = \sinh(\theta)$ and $\lambda_2 =
\cosh(\theta)$, Eq.\,(\ref{con}) can be written as
\begin{eqnarray}
\tanh(2 \theta) = \gamma /(\gamma+1),
\label{cont}
\end{eqnarray}
and in order for a solution to exist, $\gamma > -1/2$.
Equation (\ref{cont}) implies that $\theta = 
\ln(2\gamma+1)$, which can be used to express all the quantities in
terms of $\gamma$. Thus we get for the eigenvalues of $\hat H$,
\begin{equation}
{\cal E}_{n_b,n_d} - \frac \gamma 2 (N-1) = - (\gamma +1) + 
\sqrt{1 + 2 \gamma} (1 + n_b + n_d),
\label{hamapldd}
\end{equation}
where $n_b, n_d = 0,1,2, \dots$ are the eigenvalues of the number operators   
$\hat{b}^{\dagger} \hat{b}$ and $\hat{d}^{\dagger} \hat{d}$ respectively.
For $N \to \infty$, ${\cal E}_{0,0} = \gamma/2 = \epsilon_0$,
in agreement with mean field.

Having diagonalized $\hat{H}$, we can easily get the occupancy of the
$\phi_{l=\pm 1}$ states, which is
\begin{eqnarray}
\langle \hat{c}_{\pm 1}^{\dagger} \hat{c}_{\pm 1} \rangle / N 
= \lambda_1^2 / N.
\end{eqnarray}
Expanding we get that close to the transition,
\begin{eqnarray}
\frac {\langle \hat{c}_{\pm 1}^{\dagger} \hat{c}_{\pm 1} \rangle} N
\approx \frac {\sqrt 2} 8 \frac {(\gamma + 1/2)^{-1/2}} N.
\label{qu}
\end{eqnarray}
Since the occupancy of the $\phi_{\pm 1}$ states is a smooth and continuous 
function for any value of $\gamma$ when $N$ is finite, and since in the
limit $N \to \infty$, $\langle \hat{c}_{\pm 1}^{\dagger} \hat{c}_{\pm 1}
\rangle/N$ has to equal $|c_{\pm 1}|^2$ \cite{JKMR}, a simple geometric
construction shows that the value of the function in Eq.\,(\ref{qu}) at
$\gamma_+ = - 1/2 + \delta$, minus the value of $|c_{\pm 1}|^2$ at 
$\gamma_- = - 1/2 - \delta$ has to be proportional to $\delta$, and thus 
\begin{eqnarray}
  \delta = \eta N^{-2/3},
\end{eqnarray}
in agreement with Ref.\,\cite{Ueda}, where it was found numerically that 
$\eta \approx 1.077$. Therefore to leading order
\begin{eqnarray}
  |c_{\pm 1}|^2 = - \frac 4 7 (\gamma + 1/2 + \eta N^{-2/3}),
\end{eqnarray}
and 
\begin{eqnarray}
   \gamma_{\rm cr} = -1/2 - \eta N^{-2/3}.
\label{gcr}
\end{eqnarray}
   
Turning to the energy spectrum, in the lowest state $n_b = n_d = 0$,
and thus the ground state energy is 
\begin{eqnarray}
  E_0(\gamma) = - (\gamma + 1) + (2 \gamma + 1)^{1/2}
\label{eo}
\end{eqnarray}
measured from the energy of the state with uniform density,
$\gamma (N-1)/2$. Making use of Eq.\,(\ref{gcr}) we plot the function 
$E_0(\gamma+\eta N^{-2/3})$ in Fig.\,2 for $N = 100$ (top curve), 500 (middle
curve), and $N \to \infty$ (bottom curve), with $\eta = 1.077$. From 
Eq.\,(\ref{gcr}) we get that to leading order $E_0(\gamma=-1/2) = - 1/2 + 
(2 \eta)^{1/2} N^{-1/3}$.

The low-lying excited states of $\hat{H}$ lie above the lowest state by 
$E(\gamma) (n_b+n_d)$, where 
\begin{eqnarray}
  E(\gamma) = (2 \gamma + 1)^{1/2}.
\label{eoo}
\end{eqnarray}
The energy of the lowest excited state is thus $E(\gamma)$, in agreement with
Ref.\,\cite{Ueda}. Figure 3 shows $E(\gamma+\eta N^{-2/3})$ for $N = 100$ 
(top curve), 500 (middle curve), and $N \to \infty$ (bottom curve). From 
Eq.\,(\ref{gcr}), $E(\gamma=-1/2) = (2 \eta)^{1/2} N^{-1/3}$.

Finally the depletion of the condensate $\Delta N$ is, 
\begin{eqnarray}
    \Delta N = \langle \hat{c}_{-1}^{\dagger} \hat{c}_{-1} \rangle +
  \langle \hat{c}_{1}^{\dagger} \hat{c}_{1} \rangle = 2 \lambda_1^2=
  \left( \frac {\gamma + 1} {\sqrt{2 \gamma + 1}} - 1 \right).
\label{dep}
\end{eqnarray}
Figure 4 shows $\Delta N(\gamma+\eta N^{-2/3})$ for $N = 100$ (bottom curve), 
500 (middle curve), and $N \to \infty$ (top curve). According to
Eq.\,(\ref{gcr}), to leading order $\Delta N(\gamma=-1/2) = (\sqrt 2/4) 
\eta^{-1/2} N^{1/3}$.

To summarize, we examined a Bose-Einstein condensate in a toroidal
trap, with an effective attractive interaction between the atoms. Using 
a variational approach and working within an appropriately chosen set of
basis states, we demonstrated that within the mean-field approximation
for a strong enough coupling constant a condensate of uniform density 
becomes unstable against the formation of a localized state. Going
beyond the mean-field approximation, we diagonalized the Hamiltonian 
in the same truncated space just above the transition and we
calculated the energy of the low-lying states, as well as the depletion of 
the condensate. While the energy in this case is lower than that of the
mean-field, the two methods give the same results in the $N \to \infty$
limit, assumed implicitly in mean-field.

\vskip1.0pc

\noindent The author is grateful to A. D. Jackson, B. Mottelson, 
N. Papanicolaou, C. J. Pethick, and in particular to H. Smith for useful 
discussions. This work was supported by the Swedish Research Council (VR), 
and by the Swedish Foundation for Strategic Research (SSF).

\vskip 3pc

\begin{figure}
\begin{center}
\epsfig{file=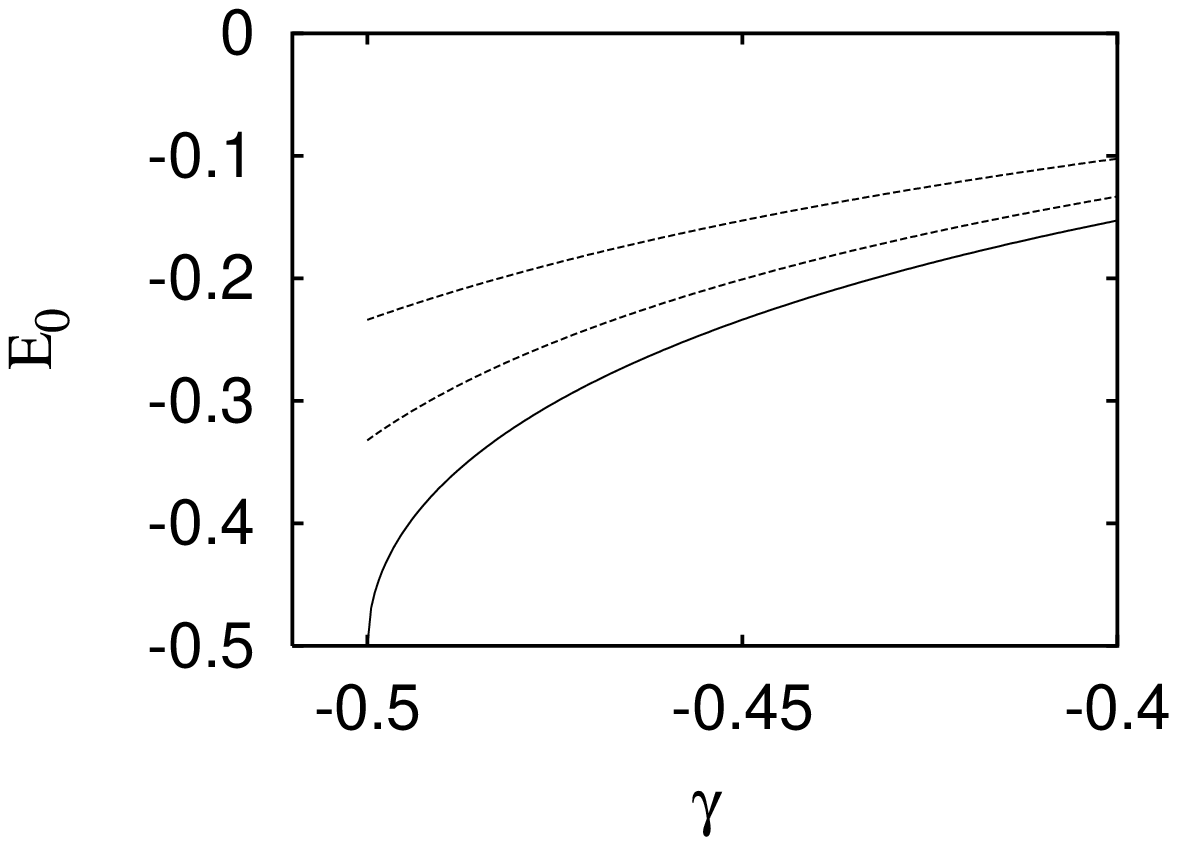,width=8.0cm,height=5.5cm,angle=0}
\vskip0.5pc
\begin{caption}
{The energy $E_0(\gamma+\eta N^{-2/3})$ of the lowest state of $\hat H$,
Eq.\,(\ref{eo}), for $N = 100$ (top curve), 500 (middle curve), and
$N \to \infty$ (bottom curve), with $\eta=1.077$.}
\end{caption}
\end{center}
\label{FIG2}
\end{figure}

\begin{figure}
\begin{center}
\epsfig{file=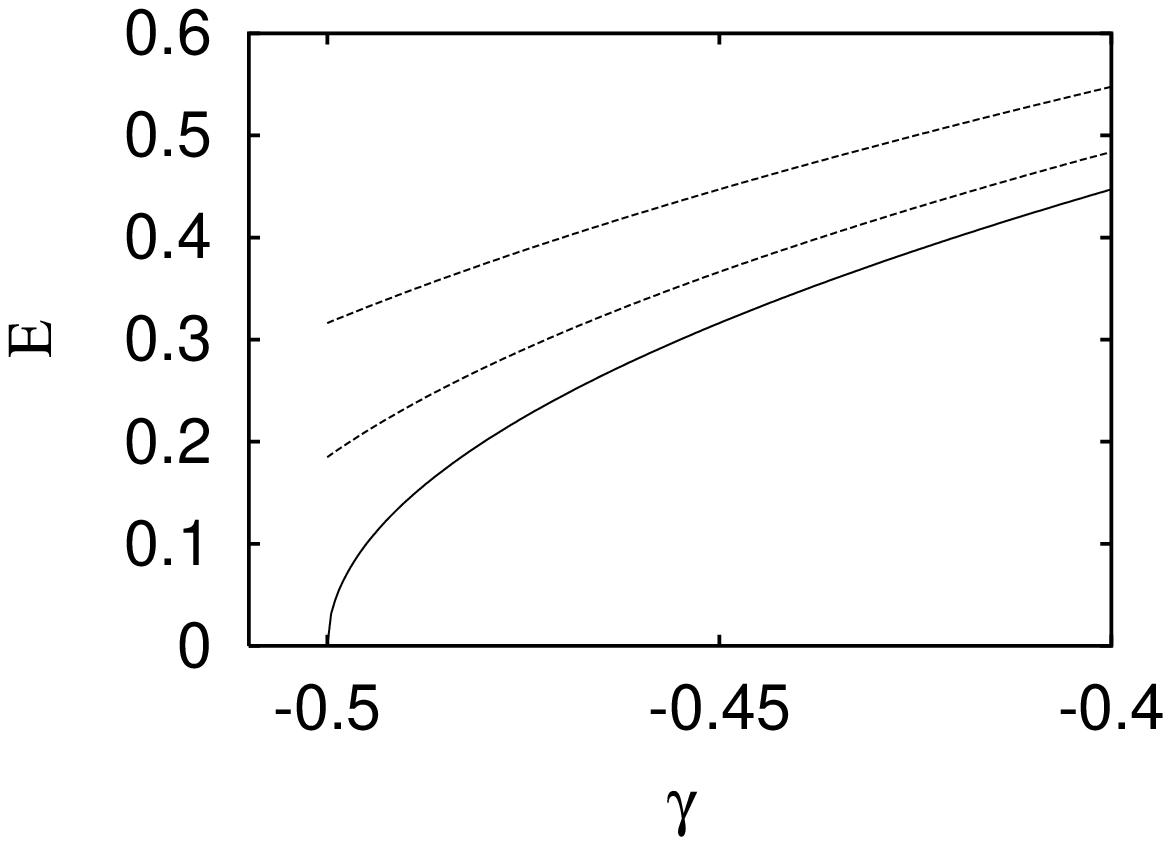,width=8.0cm,height=5.5cm,angle=0}
\vskip0.5pc
\begin{caption}
{The energy $E(\gamma+\eta N^{-2/3})$ of the first excited state of $\hat H$,
Eq.\,(\ref{eoo}), for $N = 100$ (top curve), 500 (middle curve), and
$N \to \infty$ (bottom curve), with $\eta=1.077$.}
\end{caption}
\end{center}
\label{FIG3}
\end{figure}

\begin{figure}
\begin{center}
\epsfig{file=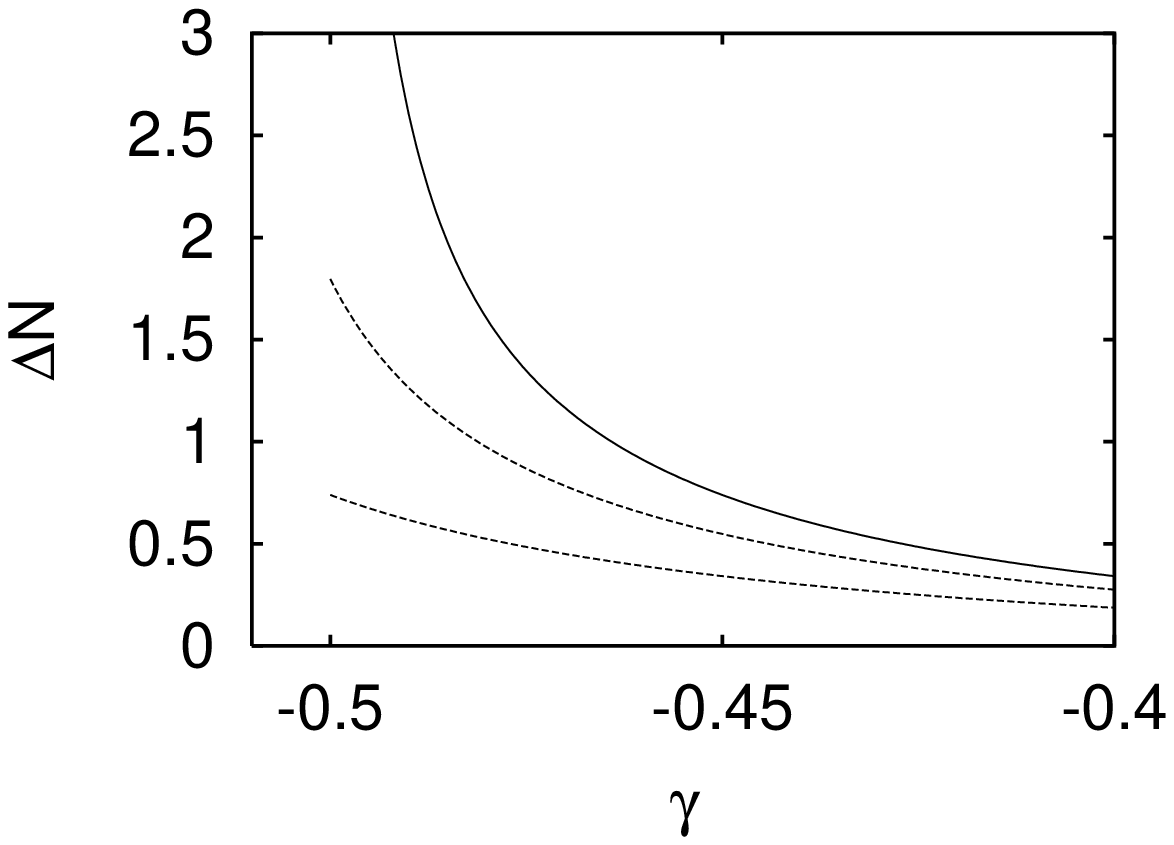,width=8.0cm,height=5.5cm,angle=0}
\vskip0.5pc
\begin{caption}
{The depletion $\Delta N(\gamma+\eta N^{-2/3})$ of the condensate, 
Eq.\,(\ref{dep}), for $N = 100$ (bottom curve), 500 (middle curve), 
and $N \to \infty$ (top curve), with $\eta=1.077$.}
\end{caption}
\end{center}
\label{FIG4}
\end{figure}


\begin{references}

\bibitem{sol1} K.~E.~Strecker, G.~B.~Partridge, A.~G.~Truscott, and
R.~G.~Hulet, Nature {\bf 417}, 150 (2002).

\bibitem{sol2} L.~Khaykovich, F.~Schreck, G.~Ferrari, T.~Bourdel,
J.~Cubizolles, L.~D.~Carr, Y.~Castin, and C.~Salomon, Science {\bf 296}, 1290
(2002).

\bibitem{Ket} S.~Inouye, M.~R.~Andrews, J.~Stenger, H.~-J.~Miesner, 
D.~M.~Stamper-Kurn, and W.~Ketterle, Nature {\bf 392}, 151 (1998).

\bibitem{refs} L.~D. Carr and Y.~Castin, e-print cond-mat/0205624;
U.~Al~Khawaja, H.~T.~C. Stoof, R.~G.~Hulet, K.~E.~Strecker, and 
G.~B.~Partridge,  e-print cond-mat/0206184; L.~Salasnich, A.~Parola, 
and L.~Reatto, e-print cond-mat/0206491.

\bibitem{Ueda} R.~Kanamoto, H.~Saito, and M.~Ueda,
e-print cond-mat/0210229.

\bibitem{Kettnew} This situation was first realized experimentally in a
Bose-Einstein condensate of $^{23}$Na atoms, as described in
A.~G\"orlitz, J.~M.~Vogels, A.~E.~Leanhardt, C.~Raman,
T.~L.~Gustavson, J.~R.~Abo-Shaeer, A.~P.~Chikkatur, S.~Gupta, S.~Inouye,
T.~P.~Rosenband, D.~E.~Pritchard, and W.~Ketterle,
Phys. Rev. Lett. {\bf 87}, 130402 (2001).

\bibitem{CCR} L.~D.~Carr, C.~W.~Clark, and W.~P.~Reinhardt,
Phys. Rev. A {\bf 62}, 063611 (2000).

\bibitem{KMP} G.~M. Kavoulakis, B.~Mottelson, and C.~J.~Pethick, 
Phys. Rev. A {\bf 62}, 063605 (2000).

\bibitem{Bog} N.~N.~Bogoliubov, J. Phys. (USSR) {\bf 11}, 23 (1947).

\bibitem{JKMR} A.~D.~Jackson, G.~M.~Kavoulakis, B.~Mottelson, and 
S.~M.~Reimann, Phys. Rev. Lett. {\bf 86}, 945 (2001).

\end{references}
\end{document}